\documentclass[conference]{IEEEtran}

\IEEEoverridecommandlockouts
\usepackage{cite}
\usepackage{amsmath,amssymb,amsfonts}
\usepackage{algorithmic}
\usepackage{graphicx}
\usepackage{textcomp}
\usepackage{xcolor}
\usepackage{url}
\usepackage{multirow}
\usepackage{xurl}

\usepackage{comment}

\usepackage[labelformat=simple]{subcaption}

\newcommand{\mlnids}{MLNIDS}
\newcommand{\thead}[1]{\textbf{#1}}

\makeatletter
\def\ps@IEEEtitlepagestyle{%
  \def\@oddfoot{\mycopyrightnotice}%
}
\def\mycopyrightnotice{%
  \begin{minipage}{\textwidth}
  \centering \scriptsize
  Copyright~\copyright~20XX IEEE. Personal use of this material is permitted. Permission from IEEE must be obtained for all other uses, in any current or future media, including reprinting/republishing this material for advertising or promotional purposes, creating new collective works, for resale or redistribution to servers or lists, or reuse of any copyrighted component of this work in other works. 
  \end{minipage}
}
\makeatother

\begin{document}
\bstctlcite{bstctl:nodash}

\title{Practical Performance of a Distributed Processing Framework for Machine-Learning-based NIDS
\thanks{This work was supported by the Hibi Science Foundation, the Naito Science \& Engineering Foundation, and Tokai Foundation for Technology.}
}

\author{\IEEEauthorblockN{Maho Kajiura}
\IEEEauthorblockA{\textit{Department of Computer Science and Engineering} \\
\textit{Toyohashi University of Technology}\\
Toyohashi, Japan \\
kmaho@dsl.cs.tut.ac.jp}
\and
\IEEEauthorblockN{Junya Nakamura}
\IEEEauthorblockA{\textit{Information and Media Center} \\
\textit{Toyohashi University of Technology}\\
Toyohashi, Japan \\
junya@imc.tut.ac.jp}
}

\maketitle

\begin{abstract}
Network Intrusion Detection Systems (NIDSs) detect intrusion attacks in network traffic. 
In particular, machine-learning-based NIDSs have attracted attention because of their high detection rates of unknown attacks. 
A distributed processing framework for machine-learning-based NIDSs employing a scalable distributed stream processing system has been proposed in the literature. 
However, its performance, when machine-learning-based classifiers are implemented has not been comprehensively evaluated. 
In this study, we implement five representative classifiers (Decision Tree, Random Forest, Naive Bayes, SVM, and kNN) based on this framework and evaluate their throughput and latency. 
By conducting the experimental measurements, we investigate the difference in the processing performance among these classifiers and the bottlenecks in the processing performance of the framework.
\end{abstract}

\section{Introduction}

A network-based intrusion detection system (NIDS) detects intrusion attacks in network traffic and notifies the network administrator.
NIDS is considered an effective defense mechanism against cyber attacks.
Traditional NIDSs detect abnormal traffic that matches intrusion attack patterns, called \emph{signatures}, stored in a system database.
However, this approach cannot detect unknown attacks because their patterns do not match the signatures.
To overcome this limitation, machine-learning-based NIDSs (\emph{{\mlnids}s}) have been proposed in recent years\cite{farhat, mahmood, De}.
{\mlnids}s detect both known and unknown attacks by building machine-learning models that include learned attack patterns based on known attacks.

Several frameworks that assist in the implementation of {\mlnids}s have been proposed\cite{Tada2019, Tun}.
These frameworks implemented all the functions necessary for {\mlnids} on scalable distributed processing systems to process network traffic efficiently.
When the network traffic increases, the {\mlnids} performance can be flexibly adapted by adding nodes.
They demonstrated the effectiveness of the framework by building {\mlnids}s and evaluating their performance in terms of throughput and latency.

However, both authors in \cite{Tada2019, Tun} did not evaluate the effectiveness of the framework when a machine-learning classifier is implemented.
Furthermore, existing {\mlnids}s also often focus on classifier performance, with only a few focusing on processing speed.
Consequently, the volume of traffic {\mlnids} can process was not investigated.
As a result, system sizing, which is necessary if the framework is actually used for {\mlnids}s, becomes difficult.

In this paper, we construct an {\mlnids} by implementing five representative classifiers based on the framework proposed in \cite{Tada2019} and evaluate their throughput and latency.
Based on this evaluation, we identify the differences in the processing performance among the classifiers and the bottlenecks in the processing performance in the framework.

The experimental results show that the processing speed and classifier performance are highly dependent on the type of classifier.
Using appropriate machine-learning algorithms, the load on the {\mlnids} can be reduced while maintaining high classifier performance.
We found that Zeek\cite{zeek}, which constructs sessions from the network traffic, Logstash\cite{logstash}, which performs the classification process using machine-learning algorithms, and Elasticsearch\cite{elasticsearch}, which stores the classification results, caused bottlenecks in the subsystems that make up the framework.

This paper is organized as follows.
Several {\mlnids} frameworks and related studies are presented in Section \ref{chapter: relation}.
Details of the existing framework proposed by Tada et al\cite{Tada2019} and the construction of {\mlnids} based on this framework is presented in Section \ref{chapter: framework}.
Our evaluation method is presented in Section \ref{chapter: measure}.
In Section \ref{chapter: result}, we present the experiment results.
Finally, Section \ref{chapter: conclusion} concludes the paper.

\section{Related Work}
\label{chapter: relation}

Various evaluation datasets for NIDSs, such as NSL-KDD\cite{nslkdd}, UNSW-NB15\cite{unswnb15}, CIC-IDS2017\cite{cicids2017}, CSE-CIC-IDS2018\cite{csecicids2018} have been proposed to assess the performance of a machine-learning algorithm in detection abnormal network traffic.
These datasets employ session-based features and have been extensively used in the evaluation of the previously proposed {\mlnids}s\cite{farhat, evalgo, comparative, kasongo, disha, mahmood}.
However, these studies evaluate only classifier performance.
To evaluate practical {\mlnids}s, it is also important to assess processing speed.

Several frameworks have been proposed for the implementation of {\mlnids}s\cite{Tun, Tada2019}.

Tun et al.~\cite{Tun} proposed a framework for {\mlnids}s that combines Apache Kafka\cite{kafka} and Spark Streaming\cite{spark}, both of which are distributed stream processing systems.
In this framework, the network traffic session is loaded into Apache Kafka via a CSV file.
Spark Streaming reads the sessions obtained from Kafka in batches of a few seconds to a few tens of seconds and classifies each session as normal or abnormal.
The authors in \cite{Tun} evaluated the processing performance of their proposed framework using UNSW-NB15.
Experimental results showed that the processing time is minimized when the batch interval of Spark Streaming is 50\,s.

Tada et al.~\cite{Tada2019} also proposed a distributed processing framework for machine-learning-based NIDS construction.
This framework consists of several stream processing subsystems and achieves high real-time performance.
Details of this framework are described in Section \ref{sec:framework-overview}.
Its basic processing performance was evaluated using UNSW-NB15.
It was demonstrated the framework can process most sessions within 500\,ms.

However, neither of the frameworks has been evaluated with machine-learning classifiers implemented.
Therefore, when building an {\mlnids} using a specific framework, the volume of network traffic that it can process is unknown.
In this paper, we implement classifiers based on the framework proposed in \cite{Tada2019} and evaluate their performance.
To investigate the practicality of {\mlnids}s, we focus on how fast the framework can process the network traffic and the load on nodes in the system.

When operating {\mlnids}, the classifier of the system needs to be updated regularly to ensure that it can detect the latest attacks.
Sato et al.~proposed a dynamic {\mlnids} operation system, which learns by acquiring normal and abnormal traffic within the target organization\cite{kobayashi}.
The system captures normal traffic from a mirror port of a router and collects abnormal traffic using a honeypot.
It extracts features from these traffic data, labels them, and rebuilds its classifier regularly.
The evaluation results show that the daily update of the classifier can keep high accuracy.

\section{Construction of {\mlnids} using a framework}
\label{chapter: framework}

In this section, we describe the construction of the {\mlnids} that will be evaluated in Section \ref{chapter: result}.
The construction of {\mlnids} is based on the distributed processing framework for machine-learning-based NIDSs proposed by Tada et al\cite{Tada2019}.
First, we present an overview of this framework in Section \ref{sec:framework-overview}; then, in Section \ref{sec:classifier-construction}, we describe the classifiers that will be implemented in the {\mlnids}.

\subsection{Framework Overview}
\label{sec:framework-overview}

\begin{figure}[tb]
    \centering
    \includegraphics[keepaspectratio, scale=0.43]{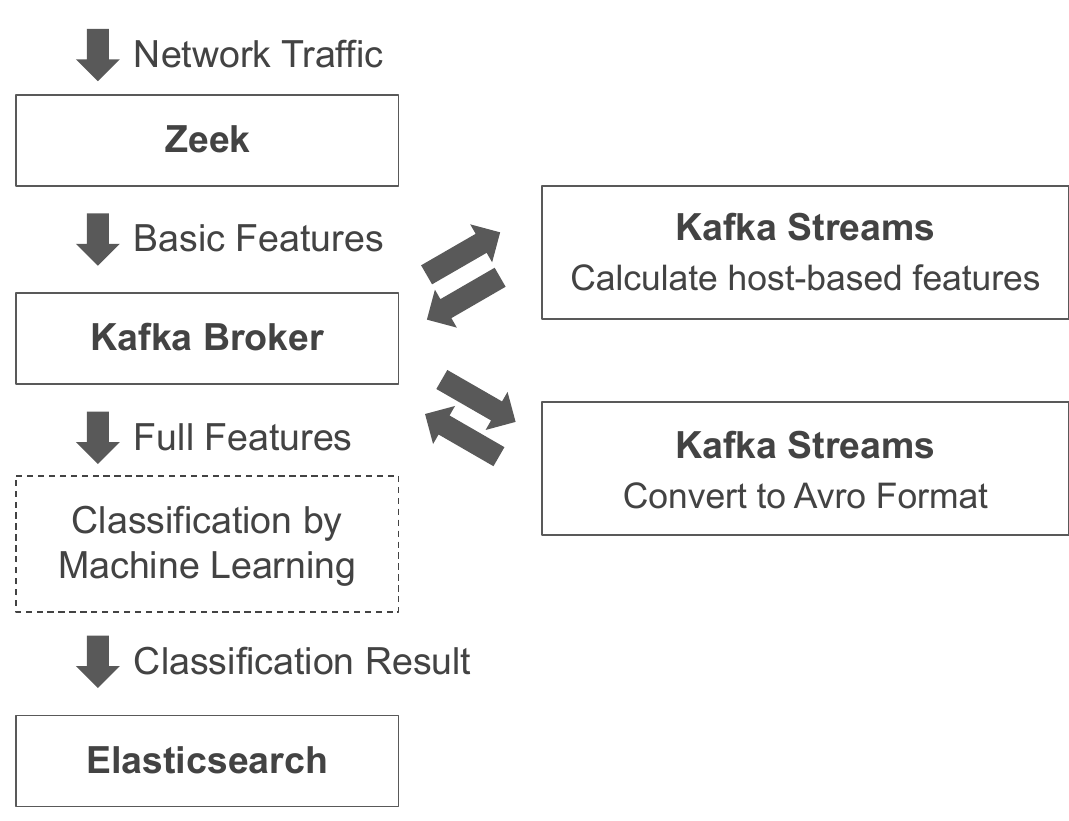}
    \caption{Process flow in the framework}
    \label{fig:framework_tada}
\end{figure}

Tada et al.~\cite{Tada2019} proposed a distributed processing framework for machine-learning-based NIDSs to facilitate the construction of practical {\mlnids}s.
Their framework consists of scalable subsystems, such as Zeek\cite{zeek}, Apache Kafka\cite{kafka}, and Elasticsearch\cite{elasticsearch}, which can improve the processing performance by increasing the number of nodes.
The framework processes the network traffic as shown in Fig.~\ref{fig:framework_tada}.
In the following, we describe the processes in each subsystem.

\begin{table}[tb]
\centering
\caption{Framework features\cite{Tada2019}}
\scalebox{1.0}{
\begin{tabular}{ll}
\hline
\thead{Name}                 & \thead{Description} \\ \hline
\multicolumn{2}{c}{\thead{Basic Features (16 types)}} \\ \hline
Timestamp & Session start time (in ms) \\
Duration                     & Session duration (in s) \\
Source IP Address            & Source IP address \\
Source Port Number           & Source port number \\
Destination IP Address       & Destination IP address \\
Destination Port Number      & Destination port number \\
Protocol & TCP, UDP, or ICMP\\
Service Type                 & HTTP, SMTP, etc. \\
Connection State & TCP connection state\\
Direction & L2L, L2R, R2L, or R2R \\
Source Packets               & Transmitted packets \\
Source Bytes                 & Transmitted bytes (payload only) \\
Source IP Bytes              & Transmitted bytes (including headers) \\
Destination Packets          & Received packets\\
Destination Bytes            & Received bytes (payload only)\\
Destination IP Bytes         & Received bytes (including headers) \\ \hline
\multicolumn{2}{c}{\thead{Host-Based Features (5 types)}} \\ \hline
DstHostCount & \#sessions with the same destination and \\
& source addresses as the current session \\
& among the last 100 sessions \\
DstHostSameSrcPortCount & \#sessions in DstHostCount whose source \\
& port is identical to the current session \\
DstHostSerrorCount & \#sessions in DstHostCount that \\
& experienced ``SYN'' errors \\
DstHostSrvCount & \#sessions with the same destination and \\
& service type as the current session among \\
& the last 100 sessions\\
DstHostSrvSerrorCount & \#sessions in DstHostSrvCount that \\
& experienced ``SYN'' errors \\
\hline
\label{tab: features}
\end{tabular}}
\end{table}

Zeek\cite{zeek} (formerly known as Bro) is an open-source network security monitoring tool, which can record, summarize, and extract network traffic.
In the framework, Zeek constructs sessions from the network traffic and prepares the basic features shown in Table \ref{tab: features} based on the session information.
These features are then sent to the Kafka Broker.

Apache Kafka\cite{kafka} is an open-source distributed event streaming platform, which provides large-scale stream data storage and distribution functions.
Apache Kafka employs the publisher-subscriber model.
Kafka Broker mediates data (messages) between publishers and subscribers.
Additionally, it manages messages in units called \emph{topic}; each topic is distributedly managed in multiple partitions.
This message management structure makes Apache Kafka highly scalable and fault-tolerant.
Apache Kafka is also equipped with a library called Kafka Streams, which helps to implement applications that process the messages stored in Apache Kafka.

The framework uses Apache Kafka as temporary storage for the features and Kafka Streams to process the features for the following two purposes:
\begin{enumerate}
    \item Based on the basic features sent by Zeek, Kafka Streams extracts the host-based features shown in Table \ref{tab: features}.
        These features are obtained from the last 100 sessions with the same destination address.
        After the extraction, Kafka Streams creates \emph{full features} by combining the basic and host-based features and stores them into a topic.
    \item In the framework, the full features are represented in their original binary format to increase its performance.
         To make the full features available from other implementations outside the framework (such as classifiers), Kafka Streams converts the full features into Apache Avro\cite{avro} format.
\end{enumerate}

The framework does not specify how {\mlnids} classifies the full features stored in the Apache Kafka Broker in Avro format.
Framework users can select any distributed stream processing framework that can use Apache Kafka as an input and Elasticsearch as an output.
For example, Kafka Streams, Apache Spark\cite{spark}, and Apache Flink\cite{flink} are potential candidates.
In this paper, we implement a classification system by combining Logstash\cite{logstash} and the well-known machine-learning library Weka\cite{weka}.
The implementation details are presented in Section \ref{sec:classifier-construction}.

Elasticsearch\cite{elasticsearch} is a distributed system that enables data perpetuation, retrieval, and stored data analysis.
It can handle various types of data, including simple data types (such as numbers and text) and complex data types (such as geospatial and structured data) in a flexible and high-speed process.
In the framework, Elasticsearch stores the full features and classification results and provides high-speed search and analysis functions for all sessions in the monitored network.

\subsection{Implementation of a Machine-Learning-based Classification}
\label{sec:classifier-construction}

Here, we describe the implementation of the classification process, which is newly implemented in this paper and incorporated into the framework.
As mentioned in Section \ref{sec:framework-overview}, we implemented the classification process by combining Logstash and Weka.

\begin{figure}[tb]
    \centering
    \includegraphics[keepaspectratio, scale=0.34]{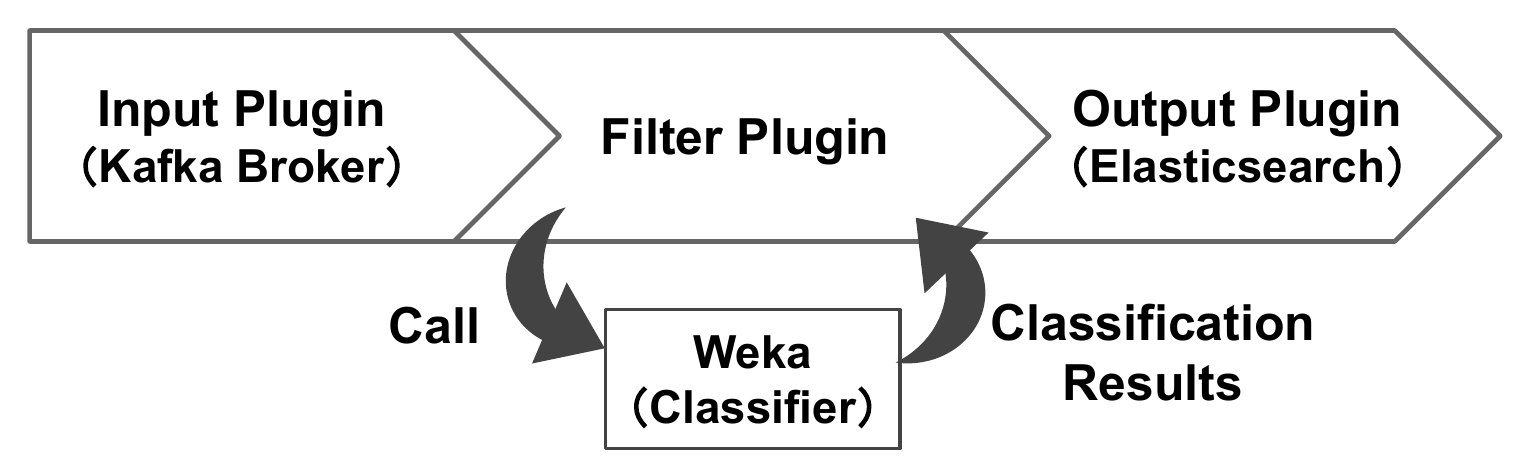}
    \caption{Logstash process flow}
    \label{fig: logstash}
\end{figure}

Logstash\cite{logstash} is a data processing pipeline developed by Elastic, Inc.; it can process data flexibly by combining different plugins at the input, filter, and output stages.
The structure of these stages in the framework is shown in Fig.~\ref{fig: logstash}.
We used the Kafka plugin and Elasticsearch plugin, which are bundled with Logstash, for the data input and output, respectively.
Additionally, we built a new filter plugin that classifies sessions by integrating with Weka.

\begin{table}[tb]
    \centering
    \caption{Changes in parameter values after conducting a parameter search}
    \scalebox{1.0}{
    \begin{tabular}{lcrlr}
        \hline
        \thead{Name} & \thead{Parameter} & \thead{Default value} & \thead{Search range} & \thead{Result} \\ 
        \hline
        \multirow{2}{*}{DT} & $C$ & 0.25 & 0.01, 0.99, 99 & 0.47 \\
         & $M$ & 2 & 1, 100, 10 & 1 \\
        \hline
        \multirow{3}{*}{RF} & $I$ & 100 & 50, 500, 10 & 100 \\
         & $N$ & 0 & 2, 5, 4 & 2 \\
         & $V$ & 1.0E-3 & 1.0E-5, 0.01, 5 & 1.0E-5 \\
        \hline
        NB & $D$ & -- & -- & enabled \\ 
        \hline
        \multirow{3}{*}{SVM} & $K$ & 2 & 0, 3, 4 & 0 \\
         & $D$ & 3 & 1, 5, 5 & 1 \\
         & $C$ & 1.0 & 0.1, 10, 100 & 8.9 \\
        \hline
        \multirow{2}{*}{kNN} & $K$ & 1 & 2, 100, 99 & 4 \\
         & $I$ & -- & -- & enabled \\
        \hline
        \label{tab: paramater}
    \end{tabular}}
\end{table}

Weka\cite{weka} is a machine-learning workbench implemented in Java; it includes various data preprocessing functions and a wide range of machine-learning algorithms. 
Weka can also be used as a Java library.
In the plugin development, we employed this feature of Weka.
We used Weka 3.8.6 in the implementation.

We employed the following five machine-learning algorithms, which are often used in {\mlnids}\cite{evalgo, comparative, farhat, kasongo}, to create the classifiers: Decision Tree (DT), Random Forest (RF), Naive Bayes (NB), SVM, and kNN\footnote{DT and kNN are implemented under the names J48 and IBk in Weka, respectively}.

In the implementation of each classifier, we performed the parameter search using the F1 score as a performance metric.
The parameters used in the parameter search are presented in Table \ref{tab: paramater}.
The search range column in Table \ref{tab: paramater} means that the parameter search is divided into the third value of steps from the first value by the second value for each parameter.
Due to page limitations, the description of each parameter is not presented here.
Readers can refer to Weka's official documentation\footnote{\url{https://weka.sourceforge.io/doc.stable-3-8/}} for the details.

Before classification, we perform the following two preprocessing stages (note that we applied the same type of preprocessing regardless of the classifier type).
First, since Timestamp, Source IP Address, and Destination IP Address in Table \ref{tab: features} are considered unimportant for classification, we remove them from the full features.
Then, by dividing each value by its maximum specification\footnote{Source Port Number, Destination Port Number, and host-based features} or that of the training data\footnote{Duration, Source Packets, Source Bytes, Source IP Bytes, Destination Packets, Destination Bytes, and Destination IP Bytes}, we normalize all the numerous features to make their values range between 0 and 1.
If a value in the test data was larger than the maximum value, we set it to 1.

\section{Evaluation method}
\label{chapter: measure}

\subsection{Dataset}

In the evaluation, we used UNSW-NB15\cite{unswnb15} as a dataset.
This dataset was recorded using realistic network traffic in the experimental network, including normal traffic and nine types of abnormal traffic (fuzzers, analysis, backdoors, DoS, exploits, generic, reconnaissance, shellcode, and worms).
It consists of the following two types of data recorded at different periods:
\begin{itemize}
    \item 1 abnormal traffic per second (January 22, 2015)
    \item 10 abnormal traffics per second (February 17, 2015)
\end{itemize}
We used the first type of data to train the classifier and the second one to evaluate the classification performance and measure the processing speed.

\begin{table}[tb]
    \centering
    \caption{Number of sessions in the dataset}
    \label{tab: dataset}
    \scalebox{1.0}{
    \begin{tabular}{llccc}
        \hline
        \multirow{2}{*}{\thead{Date}} & \multirow{2}{*}{\thead{Purpose}} & \thead{Total} & \thead{Normal} & \thead{Abnormal} \\
         & & \thead{sessions} & \thead{sessions} & \thead{sessions} \\
        \hline
        January 22, 2015 & training & 1,028,319 & 1,024,430 & \multicolumn{1}{r}{3,889} \\
        February 17, 2015 & evaluation & 1,030,117 & 1,010,661 & \multicolumn{1}{r}{19,456} \\
        \hline
    \end{tabular}}
\end{table}

The features of UNSW-NB15 cannot be used directly in this framework because they are different from those used in this paper (Table \ref{tab: features}).
Therefore, we converted the features as follows.
First, we read UNSW-NB15 PCAP data and generated the features of the framework using Zeek.
Then, we assigned the correct labels to the generated features based on the UNSW-NB15 dataset.
The total number of sessions after the conversion is presented in Table \ref{tab: dataset}.

Additionally, we performed downsampling to balance the number of normal and abnormal sessions during classifier training and evaluation of its classification performance.
This is because a large imbalance in the number of samples between classes reduces the sensitivity of the model to minority classes\cite{abdulhammed}.

\subsection{Experimental Environment}

\begin{table*}[tb]
\centering
\caption{Kubernetes cluster configuration}
\label{tab: k8s_cluster}
\scalebox{1.0}{
\begin{tabular}{lrllr}
\hline
\thead{Subsystem} & \thead{Version}  & \thead{Pod name} & \thead{Pod role} & \thead{Deployment Node} \\
\hline
Zeek & 4.0.4 & \texttt{zeek} & Run Zeek & 1 \\ \hline
Kafka Broker & 3.4.0 & \texttt{kafka-broker} & Run Kafka Broker & 2 \\ \hline
\multirow{5}{*}{Kafka Streams} & \multirow{5}{*}{2.0.0} & \texttt{streams-fe} & Calculate host-based features & 4\\
 &  & \texttt{streams-fc-0} & Convert data from the original format to the Avro format & 5\\
 &  & \texttt{streams-fc-1} & Convert data from the original format to the Avro format & 5\\
 &  & \texttt{streams-fc-2} & Convert data from the original format to the Avro format & 5\\
 &  & \texttt{streams-fc-3} & Convert data from the original format to the Avro format & 5\\
\hline
\multirow{2}{*}{Logstash} & \multirow{2}{*}{8.7.0}  & \texttt{logstash-0} & Classify sessions as normal/abnormal traffic & 6 \\
 &  & \texttt{logstash-1} & Classify sessions as normal/abnormal traffic & 7 \\
\hline
\multirow{8}{*}{Elasticsearch} & \multirow{8}{*}{8.6.2} & \texttt{master-0} & Manage Elasticsearch clusters & 3 \\
 & & \texttt{master-1} & Manage Elasticsearch clusters & 3 \\
 & & \texttt{coordinating-0} & Receive client requests & 3 \\
 & & \texttt{coordinating-1} & Receive client requests & 3 \\
 & & \texttt{ingest-0} & Process data preprocessing and transformation & 8 \\
 & & \texttt{ingest-1} & Process data preprocessing and transformation & 8 \\
 & & \texttt{data-0} & Store data & 9 \\
 & & \texttt{data-1} & Store data & 9 \\
\hline
\end{tabular}}
\end{table*}

\begin{table}[tb]
\centering
\caption{Specifications of each node}
\label{tab: machines}
\scalebox{1.0}{
\begin{tabular}{cll}
\hline
\thead{Node number} & \thead{CPU} & \thead{Memory} \\
\hline
\multirow{2}{*}{1} & Xeon Silver 4214R 2.4 GHz & DDR4-2933 \\
 & 12 Core 24 Thread & ECC 16 GB \\ 
\hline
\multirow{2}{*}{2, 3} & Xeon E-2136 3.3 GHz & DDR4-2666 \\
 & 6 Core 12 Thread & ECC 16 GB \\ 
\hline
4, 5, 6 & Xeon E3-1220v5 3.0 GHz & DDR4-2133 \\
7, 8, 9 & 4 Core 4 Thread & ECC 16 GB \\ 
\hline
\end{tabular}}
\end{table}

We conducted experiments on a Kubernetes\cite{k8s} cluster composed of nine nodes running Ubuntu 20.04.
Kubernetes is a system for automating deployment, scaling, and management of containerized applications and manages container applications in a unit called a \emph{pod}, which consists of one or more containers.
The configuration of the Kubernetes cluster and the specifications of each node are presented in Tables \ref{tab: k8s_cluster} and \ref{tab: machines}, respectively.
In this setting, we used four pods to convert features from the original binary format into the Avro format using Kafka Streams in parallel to improve efficiency.
Similarly, we used two Logstash pods to accelerate the classification performance.
In addition, we deployed two pods for each Elasticsearch service to improve fault tolerance.

\subsection{Performance Metrics}

In the experiment, we measured the processing speed of the framework in terms of throughput and latency as follows.
 
The throughput calculation was based on the number of sessions inserted into Elasticsearch per second.
First, the measurement period was divided into 30\,s intervals and the average number of sessions per second was calculated for each interval.
Next, the maximum of the calculated average for each interval was considered as the throughput under the experimental conditions.
 
The latency calculation was based on the time between the creation of a session by Zeek and its insertion into Elasticsearch.
First, the measurement period was divided into 10\,s intervals, and the average latency was calculated for each interval.
The median of the average latency for each interval is the latency under the experimental conditions.

We measured the classifier performance using the F1 score, which is the harmonic mean of Precision and Recall.
These metrics are expressed as follows:
\begin{displaymath}
    \mathit{Precision} = \frac{\mathit{TP}}{\mathit{TP}+\mathit{FP}},\  \mathit{Recall} = \frac{\mathit{TP}}{\mathit{TP}+\mathit{FN}}
\end{displaymath}
\begin{displaymath}
    \mathit{F1\ score} = \frac{2 \times \mathit{Precision} \times \mathit{Recall}}{\mathit{Precision}+\mathit{Recall}}
\end{displaymath}
Here, \textit{TP}, \textit{FP}, and \textit{FN} are true positive rate, false positive rate, and false negative rate, respectively.

\subsection{Performance Limits in the Experimental Environment}
\label{sec: measure_MaxSpeed}

As a preliminary experiment, we measured the maximum traffic that can be loaded on the framework in our experimental environment.
First, we measured the maximum number of sessions that one Zeek process can handle.
We ran a Zeek pod on Node 1 shown in Table \ref{tab: machines}.
We used the PCAP data of UNSW-NB15 as the network traffic and gradually increased the data transmission rate to the Zeek pod using the \verb,tcpreplay, command.
As a result, at a data transmission rate of 1\,Gbps (about 2,700 sessions per second), the CPU usage of the Zeek pod reached 100\%, which was the upper limit of processing speed using one Zeek process.

Since Zeek is not designed to be multithreaded, in a normal configuration, Zeek cannot process sessions with a CPU usage rate greater than 100\% (i.e., 1\,Gbps in the current configuration).
Therefore, we run multiple Zeek processes so that these processes can handle the same network traffic.
This can cumulatively increase the network traffic processed by the framework.
In this way, we increased the number of Zeek processes incrementally.
When running nine Zeek processes, the CPU usage of all CPU cores on Node 1 reached 100\%, and no more sessions could be processed, i.e., the maximum load that can be applied to the framework in this experimental environment is 9\,Gbps (about 24,300 sessions per second).
Therefore, we should note that the processing speed of each classifier in Section \ref{chapter: result} is limited by this speed.

\section{Experimental Results}
\label{chapter: result}

\subsection{Classifier Performance}
\label{sec: F1}

\begin{table}[tb]
\centering
\caption{F1 score of each classifier}
\label{tab: F1}
\scalebox{1.0}{
\begin{tabular}{lrrrrr}
\hline
\thead{Parameter}& \thead{DT} & \thead{RF} & \thead{NB} & \thead{SVM} & \thead{kNN} \\
\hline
Default value & 0.967 & 0.974 & 0.599 & 0.753 & 0.866 \\
Value after PS & 0.967 & 0.972 & 0.964 & 0.855 & 0.873 \\ \hline
\end{tabular}}
\end{table}

First, we compared the classification performance of each machine-learning algorithm using its default parameters with the optimal parameters found by performing a parameter search.
The F1 scores of the classifiers are presented in Table~\ref{tab: F1}.

A comparison of the classifiers based on their F1 scores shows that  RF achieves the highest score (0.974), followed by DT (0.967), NB (0.964), kNN (0.873), and SVM (0.855).
A similar trend regarding the superiority of RF and DT is observed in the experimental results reported in \cite{farhat}.

The F1 score of RF slightly decreased after performing the parameter search, while the F1 scores of the other classifiers were improved or remained unchanged.
We attribute the decrease in the F1 score of RF to the overfitting caused by the low value of parameter $V$.
This parameter controls the time a numeric feature is split in a tree.
A low value of $V$ is likely to generate a complex and detailed tree.

\subsection{Maximum Processing Speed}
\label{sec: speed}

\begin{figure}[tb]
    \centering
    \includegraphics[keepaspectratio, width=\linewidth]{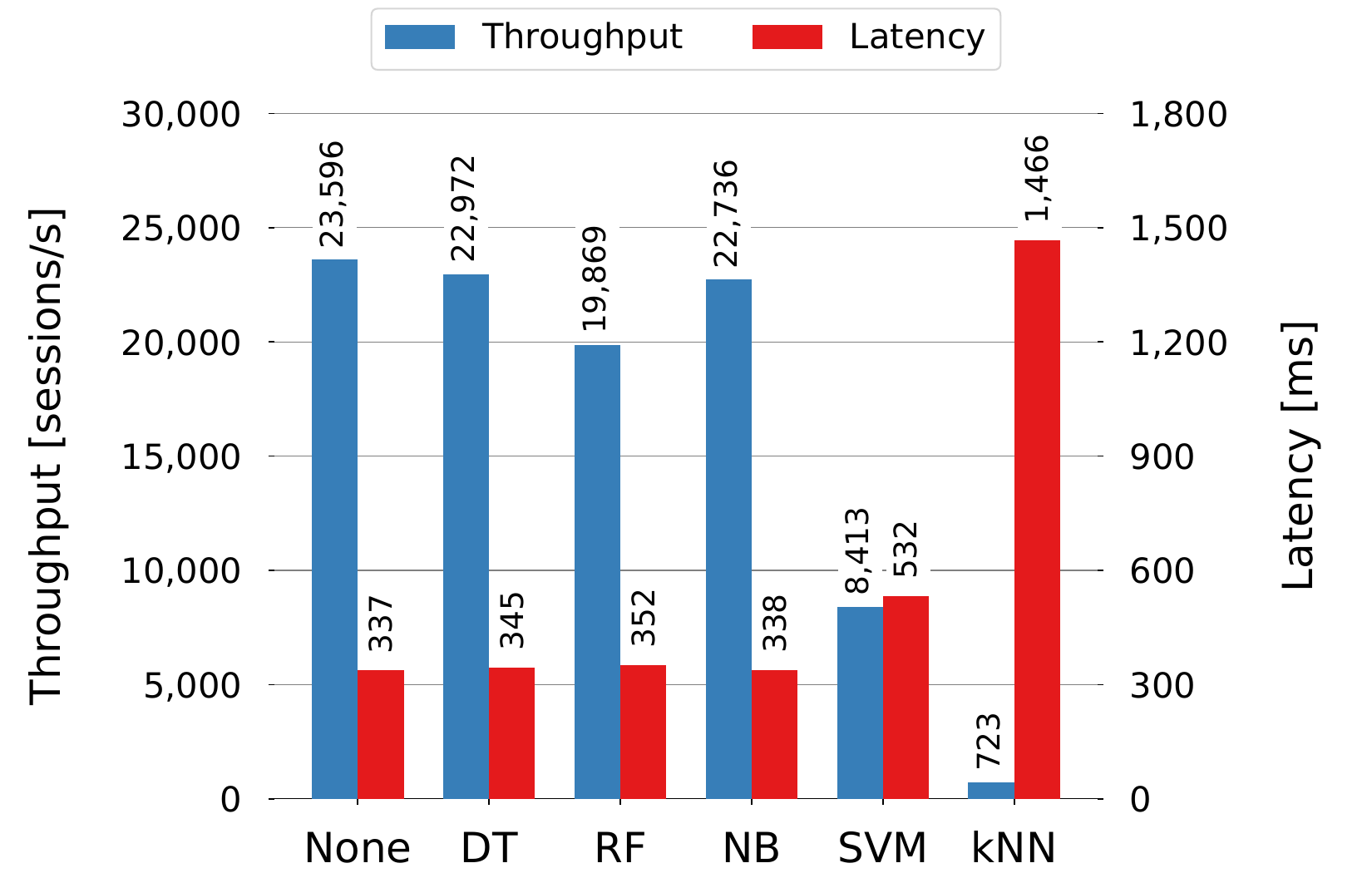}
    \caption{Throughput and latency achieved each classifier in the framework}
    \label{fig: max_speed}
\end{figure}

We evaluated the throughput and latency of the classifiers employed in the framework.
For each machine-learning algorithm, we employed the classifier with the highest F1 score in Section \ref{sec: F1}.
As a baseline, we also measured the processing speed without a classifier (i.e., in Fig.~\ref{fig: logstash}, the filter plugin does not process anything).
The throughput and latency of each classifier are shown in Fig.~\ref{fig: max_speed}.
Each value is the average of three measurement values obtained under the same conditions.

The throughput results show that DT and NB achieve the fastest processing speeds.
Those two classifiers were able to process classifications at a speed of 22,972 sessions per second and 22,736 sessions per second, respectively.
These results are almost equal to the results obtained without employing a classifier (23,596 sessions per second).
The throughputs were measured by loading Zeek with 9.0\,Gbps network traffic, which is the upper limit of the environment monitored in Section \ref{sec: measure_MaxSpeed}.
Therefore, it seems that DT and NB can process network traffic of over 9.0\,Gbps.
The next fastest throughput was achieved by RF with 19,869 sessions per second (equivalent to 7.5 Gbps), followed by SVM with 8,413 sessions per second (equivalent to 3.0 Gbps) and kNN with 723 sessions per second (equivalent to 250 Mbps).
 
Regarding latency, NB, DT, and RF achieved about 340--350\,ms, followed by SVM with 530\,ms and kNN with 1.5\,s.

The results show a significant difference in the throughput and latency among classifiers.
For example, DT and NB can process about 9.0 Gbps, whereas RF can only process about 7.5 Gbps.
In contrast, the classification performance (F1 score) of DT, NB, and RF is almost the same (around 0.97); however, the acceptable network traffic is quite different.
This indicates that the classification performance can be improved using fewer computational resources by selecting the appropriate machine-learning algorithm.
While DT, RF, and NB increase latency by a few milliseconds to a few tens of milliseconds compared to not using any classifiers, this small overhead is acceptable for real-time performance.

We recommend the following classifier selections under different traffic conditions.
First, if the traffic is less than 7.5\,Gbps, RF should be used because it showed the best classifier performance among the five classifiers.
If the traffic exceeds 7.5\,Gbps, we recommend using DT or NB since RF cannot process traffic faster than 7.5\,Gbps, and DT and NB had the second-best classifier performance.

\subsection{CPU Usage of Each Subsystem}

\begin{figure}[tb]
    \centering
    \begin{minipage}[b]{0.95\linewidth}
        \centering
        \includegraphics[keepaspectratio, scale=0.16]{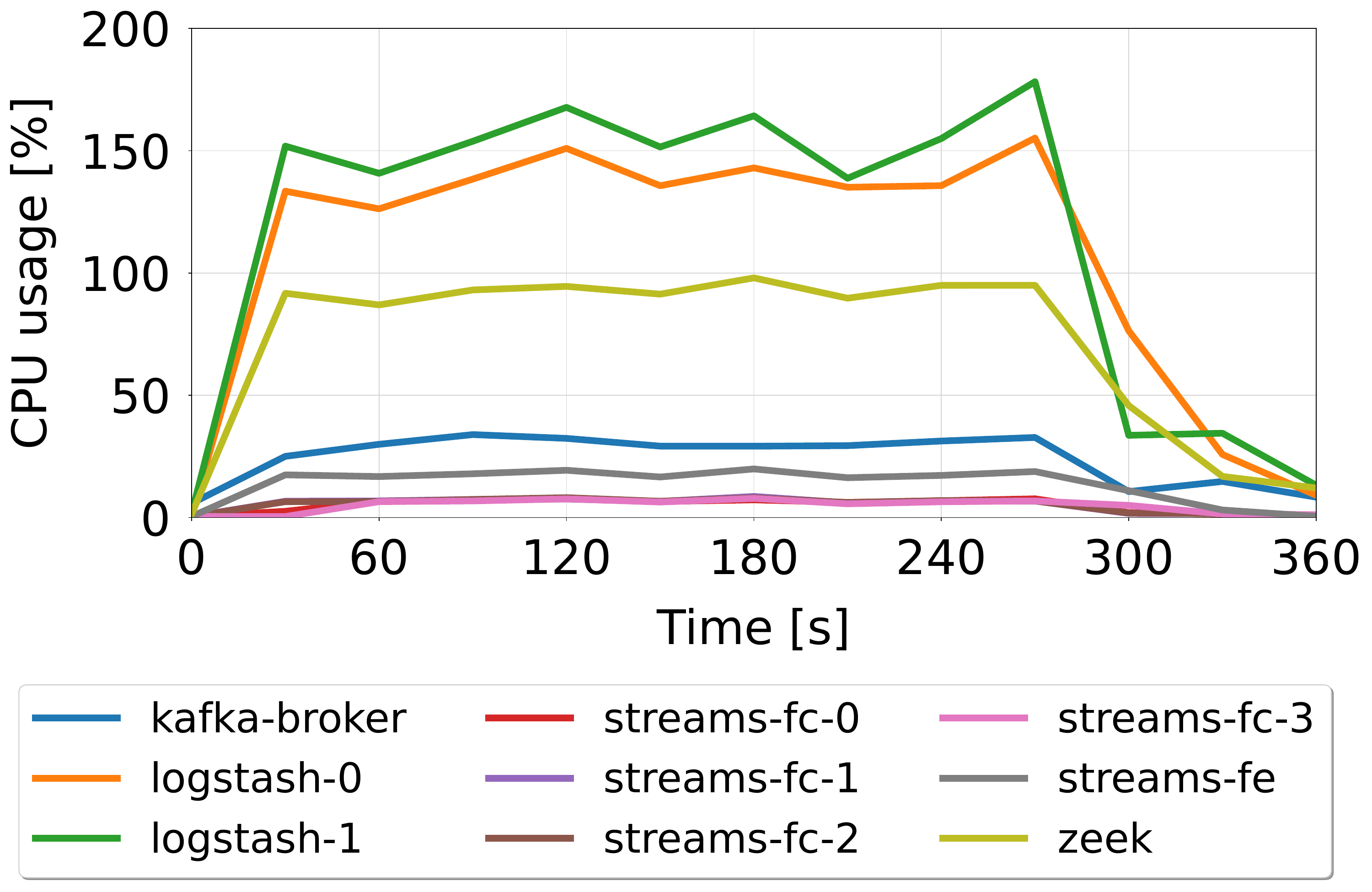}
        \subcaption{Pods other than Elasticsearch}
        \label{subfig: usage_others}
    \end{minipage}
    \begin{minipage}[b]{0.95\linewidth}
        \centering
        \includegraphics[keepaspectratio, scale=0.16]{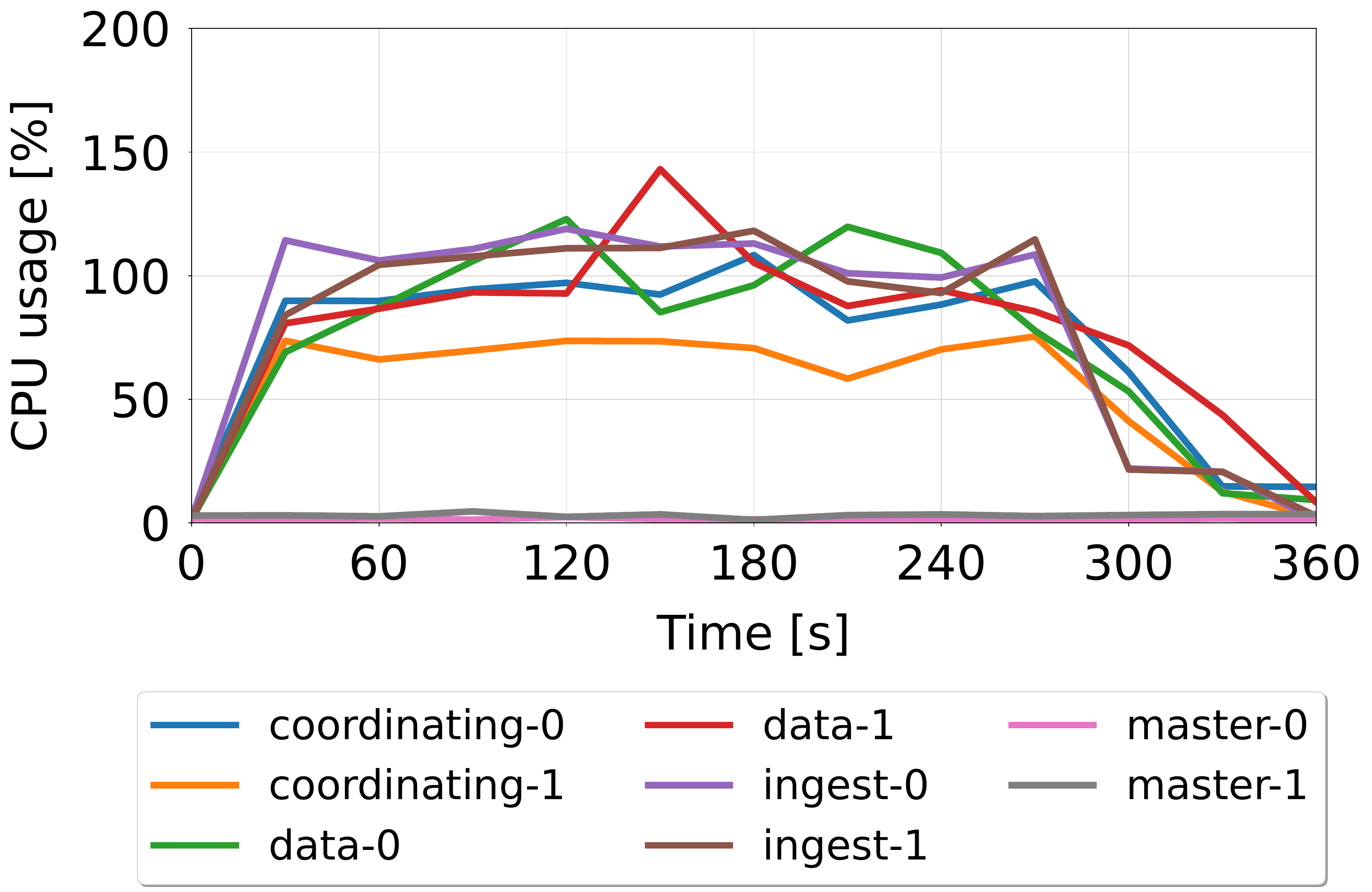}
        \subcaption{Elasticsearch Pods}
        \label{subfig: usage_es}
    \end{minipage}
    \caption{CPU usage of each pod}
    \label{fig: cpu_usage}
\end{figure}

We compared the CPU usage of each pod to investigate the processing bottleneck in the framework.
As an example, Fig.~\ref{fig: cpu_usage} shows the CPU usage of each pod when processing 7.5 Gbps of network traffic using RF.

Fig.~\ref{subfig: usage_others} shows that the CPU usage of the Logstash pods (\verb,logstash-0, and \verb,logstash-1,) and \verb,zeek, pod is high (over 130\% and about 100\%, respectively).
In contrast, the CPU usage of \verb,kafka-broker, pod, \verb,streams-fe, pod, and all four \verb,streams-fc, pods is low (about 30\%, about 20\%, and about 10\%, respectively).
Fig.~\ref{subfig: usage_es} shows that the CPU usage of five Elasticsearch pods (\verb,coordinating-0,, \verb,data-0,, \verb,data-1,, \verb,ingest-0, and \verb,ingest-1,) exceeds 100\%. 
The CPU usage of \verb,coordinating-1, pod exceeds 58\%.
On the other hand, the CPU usage of \verb,master-0, and \verb,master-1, pods is less than 5\%.

The above results show that the subsystems with the highest CPU usage are Zeek and Logstash, as well as Elasticsearch pods that are used for data, ingesting, and coordinating.
As discussed in Section \ref{sec: measure_MaxSpeed}, the maximum CPU usage of Zeek for one process (i.e., one pod) is 100\% (it can process about 2,700 sessions per second).
If the number of sessions per second exceeds 2,700, we need to run multiple Zeek processes and distribute the network traffic across the processes. 
Logstash and Elasticsearch are also likely to be overloaded because they are responsible for classifying sessions and storing the classification results, respectively.
The performance of Logstash and Elasticsearch needs to be improved by allocating pods to high-performance nodes or by processing them in parallel on multiple pods.

\section{Conclusion}
\label{chapter: conclusion}

In practical applications of machine-learning-based network intrusion detection systems ({\mlnids}s), both classifier performance and processing speed are important.
However, these aspects have received limited attention.
In this study, we implemented several classifiers based on the distributed processing framework for {\mlnids}s proposed in a previous study and evaluated their performance.
The experimental results showed that by selecting appropriate machine-learning algorithms, the load can be reduced while maintaining the {\mlnids} classifier performance.
The CPU usage analysis shows that the Zeek, Logstash, and Elasticsearch processes can cause bottlenecks.

In a future study, we will reconstruct the experimental environment to evaluate the performance of DT and NB in the framework with over 9\,Gbps of network traffic.
Additionally, we will evaluate the performance of deep-learning-based classifiers.
It is also important to evaluate the stability and fault tolerance of {\mlnids}s when operating the system for a long time.
Furthermore, we consider the periodic updating of the classifiers to maintain their classifier performance.

\bibliographystyle{IEEEtran}
\bibliography{ref}

\end{document}